\documentclass[twocolumn,letterpaper,aps,prc,longbibliography,superscriptaddress,nofootinbib,floatfix]{revtex4-2}

\usepackage{graphicx}   
\usepackage{tabularx}
\usepackage{amsmath}
\usepackage[normalem]{ulem}
\usepackage{xspace}	

\newcommand{\pt}{\mbox{$p_T$}\xspace}
\newcommand{\Gevc}{\mbox{GeV/$c$}\xspace}

\newcommand{\Mevcc}{\mbox{MeV/c$^2$}\xspace}
\newcommand{\rxa}{\mbox{$R_{xA}$}\xspace}
\newcommand{\Npart}{\mbox{$\langle N_{\rm part} \rangle$}\xspace}
\newcommand{\Ncoll}{\mbox{$\langle N_{\rm coll} \rangle$}\xspace}

\newcommand{\sqsn}{\mbox{$\sqrt{s_{_{NN}}}$}\xspace}
\newcommand{\sqs}{\mbox{$\sqrt{s}$}\xspace}
\newcommand{\sqsntwo}{\mbox{$\sqrt{s_{_{NN}}}=200$~GeV}\xspace}

\newcommand{\pp}{\mbox{$p$$+$$p$}\xspace}
\newcommand{\dau}{\mbox{$d$$+$Au}\xspace}

\newcommand{\pau}{\mbox{$p$$+$Au}\xspace}
\newcommand{\pal}{\mbox{$p$$+$Al}\xspace}
\newcommand{\heau}{\mbox{$^3$He$+$Au}\xspace}
\newcommand{\pdheau}{\mbox{$p/d/^3$He$+$Au}\xspace}\newcommand{\auau}{\mbox{Au$+$Au}\xspace}
\newcommand{\cucu}{\mbox{Cu$+$Cu}\xspace}

\newcommand{\vphi}{\mbox{$\phi$}\xspace}
\newcommand{\pio}{\mbox{$\pi^0$}\xspace}
\newcommand{\et}{\mbox{$\eta$}\xspace}

\newcommand{\Jpsi}{\mbox{$J/\psi$}\xspace}
\newcommand{\PsiSS}{\mbox{$\psi(2S)$}\xspace}

\begin{document}

\title{ Study of $\phi$ meson production in $p$$+$Al, $p$$+$Au, 
$d$$+$Au, and $^3$He$+$Au collisions at $\sqrt{s_{_{NN}}}=200$ GeV }

\newcommand{\abilene}{Abilene Christian University, Abilene, Texas 79699, USA}
\newcommand{\augie}{Department of Physics, Augustana University, Sioux Falls, South Dakota 57197, USA}
\newcommand{\banaras}{Department of Physics, Banaras Hindu University, Varanasi 221005, India}
\newcommand{\barc}{Bhabha Atomic Research Centre, Bombay 400 085, India}
\newcommand{\baruch}{Baruch College, City University of New York, New York, New York, 10010 USA}
\newcommand{\bnlcoll}{Collider-Accelerator Department, Brookhaven National Laboratory, Upton, New York 11973-5000, USA}
\newcommand{\bnlphys}{Physics Department, Brookhaven National Laboratory, Upton, New York 11973-5000, USA}
\newcommand{\caucr}{University of California-Riverside, Riverside, California 92521, USA}
\newcommand{\charlesczech}{Charles University, Ovocn\'{y} trh 5, Praha 1, 116 36, Prague, Czech Republic}
\newcommand{\ciae}{Science and Technology on Nuclear Data Laboratory, China Institute of Atomic Energy, Beijing 102413, People's Republic of China}
\newcommand{\cns}{Center for Nuclear Study, Graduate School of Science, University of Tokyo, 7-3-1 Hongo, Bunkyo, Tokyo 113-0033, Japan}
\newcommand{\colorado}{University of Colorado, Boulder, Colorado 80309, USA}
\newcommand{\columbia}{Columbia University, New York, New York 10027 and Nevis Laboratories, Irvington, New York 10533, USA}
\newcommand{\czechtech}{Czech Technical University, Zikova 4, 166 36 Prague 6, Czech Republic}
\newcommand{\debrecen}{Debrecen University, H-4010 Debrecen, Egyetem t{\'e}r 1, Hungary}
\newcommand{\elte}{ELTE, E{\"o}tv{\"o}s Lor{\'a}nd University, H-1117 Budapest, P{\'a}zm{\'a}ny P.~s.~1/A, Hungary}
\newcommand{\ewha}{Ewha Womans University, Seoul 120-750, Korea}
\newcommand{\famu}{Florida A\&M University, Tallahassee, FL 32307, USA}
\newcommand{\fsu}{Florida State University, Tallahassee, Florida 32306, USA}
\newcommand{\gsu}{Georgia State University, Atlanta, Georgia 30303, USA}
\newcommand{\hiroshima}{Hiroshima University, Kagamiyama, Higashi-Hiroshima 739-8526, Japan}
\newcommand{\howard}{Department of Physics and Astronomy, Howard University, Washington, DC 20059, USA}
\newcommand{\ihepprot}{IHEP Protvino, State Research Center of Russian Federation, Institute for High Energy Physics, Protvino, 142281, Russia}
\newcommand{\illuiuc}{University of Illinois at Urbana-Champaign, Urbana, Illinois 61801, USA}
\newcommand{\inrras}{Institute for Nuclear Research of the Russian Academy of Sciences, prospekt 60-letiya Oktyabrya 7a, Moscow 117312, Russia}
\newcommand{\instpasczech}{Institute of Physics, Academy of Sciences of the Czech Republic, Na Slovance 2, 182 21 Prague 8, Czech Republic}
\newcommand{\isu}{Iowa State University, Ames, Iowa 50011, USA}
\newcommand{\jaea}{Advanced Science Research Center, Japan Atomic Energy Agency, 2-4 Shirakata Shirane, Tokai-mura, Naka-gun, Ibaraki-ken 319-1195, Japan}
\newcommand{\jeonbuk}{Jeonbuk National University, Jeonju, 54896, Korea}
\newcommand{\jyvaskyla}{Helsinki Institute of Physics and University of Jyv{\"a}skyl{\"a}, P.O.Box 35, FI-40014 Jyv{\"a}skyl{\"a}, Finland}
\newcommand{\kek}{KEK, High Energy Accelerator Research Organization, Tsukuba, Ibaraki 305-0801, Japan}
\newcommand{\korea}{Korea University, Seoul 02841, Korea}
\newcommand{\kurchatov}{National Research Center ``Kurchatov Institute", Moscow, 123098 Russia}
\newcommand{\kyoto}{Kyoto University, Kyoto 606-8502, Japan}
\newcommand{\lawllnl}{Lawrence Livermore National Laboratory, Livermore, California 94550, USA}
\newcommand{\losalamos}{Los Alamos National Laboratory, Los Alamos, New Mexico 87545, USA}
\newcommand{\lund}{Department of Physics, Lund University, Box 118, SE-221 00 Lund, Sweden}
\newcommand{\lyon}{IPNL, CNRS/IN2P3, Univ Lyon, Université Lyon 1, F-69622, Villeurbanne, France}
\newcommand{\maryland}{University of Maryland, College Park, Maryland 20742, USA}
\newcommand{\mass}{Department of Physics, University of Massachusetts, Amherst, Massachusetts 01003-9337, USA}
\newcommand{\mate}{MATE, Laboratory of Femtoscopy, K\'aroly R\'obert Campus, Gy\"ongy\"os, Hungary}
\newcommand{\michigan}{Department of Physics, University of Michigan, Ann Arbor, Michigan 48109-1040, USA}
\newcommand{\miss}{Mississippi State University, Mississippi State, Mississippi 39762, USA}
\newcommand{\muhlenberg}{Muhlenberg College, Allentown, Pennsylvania 18104-5586, USA}
\newcommand{\nara}{Nara Women's University, Kita-uoya Nishi-machi Nara 630-8506, Japan}
\newcommand{\natmephi}{National Research Nuclear University, MEPhI, Moscow Engineering Physics Institute, Moscow, 115409, Russia}
\newcommand{\newmex}{University of New Mexico, Albuquerque, New Mexico 87131, USA}
\newcommand{\nmsu}{New Mexico State University, Las Cruces, New Mexico 88003, USA}
\newcommand{\northcg}{Physics and Astronomy Department, University of North Carolina at Greensboro, Greensboro, North Carolina 27412, USA}
\newcommand{\ohio}{Department of Physics and Astronomy, Ohio University, Athens, Ohio 45701, USA}
\newcommand{\ornl}{Oak Ridge National Laboratory, Oak Ridge, Tennessee 37831, USA}
\newcommand{\orsay}{IPN-Orsay, Univ.~Paris-Sud, CNRS/IN2P3, Universit\'e Paris-Saclay, BP1, F-91406, Orsay, France}
\newcommand{\peking}{Peking University, Beijing 100871, People's Republic of China}
\newcommand{\pnpi}{PNPI, Petersburg Nuclear Physics Institute, Gatchina, Leningrad region, 188300, Russia}
\newcommand{\pusan}{Pusan National University, Pusan 46241, Korea}
\newcommand{\riken}{RIKEN Nishina Center for Accelerator-Based Science, Wako, Saitama 351-0198, Japan}
\newcommand{\rikjrbrc}{RIKEN BNL Research Center, Brookhaven National Laboratory, Upton, New York 11973-5000, USA}
\newcommand{\rikkyo}{Physics Department, Rikkyo University, 3-34-1 Nishi-Ikebukuro, Toshima, Tokyo 171-8501, Japan}
\newcommand{\saispbstu}{Saint Petersburg State Polytechnic University, St.~Petersburg, 195251 Russia}
\newcommand{\seoulnat}{Department of Physics and Astronomy, Seoul National University, Seoul 151-742, Korea}
\newcommand{\stonybrkc}{Chemistry Department, Stony Brook University, SUNY, Stony Brook, New York 11794-3400, USA}
\newcommand{\stonycrkp}{Department of Physics and Astronomy, Stony Brook University, SUNY, Stony Brook, New York 11794-3800, USA}
\newcommand{\tenn}{University of Tennessee, Knoxville, Tennessee 37996, USA}
\newcommand{\texsu}{Texas Southern University, Houston, TX 77004, USA}
\newcommand{\titech}{Department of Physics, Tokyo Institute of Technology, Oh-okayama, Meguro, Tokyo 152-8551, Japan}
\newcommand{\tsukuba}{Tomonaga Center for the History of the Universe, University of Tsukuba, Tsukuba, Ibaraki 305, Japan}
\newcommand{\vandy}{Vanderbilt University, Nashville, Tennessee 37235, USA}
\newcommand{\weizmann}{Weizmann Institute, Rehovot 76100, Israel}
\newcommand{\wigner}{Institute for Particle and Nuclear Physics, Wigner Research Centre for Physics, Hungarian Academy of Sciences (Wigner RCP, RMKI) H-1525 Budapest 114, POBox 49, Budapest, Hungary}
\newcommand{\yonsei}{Yonsei University, IPAP, Seoul 120-749, Korea}
\newcommand{\zagreb}{Department of Physics, Faculty of Science, University of Zagreb, Bijeni\v{c}ka c.~32 HR-10002 Zagreb, Croatia}
\newcommand{\zambia}{Department of Physics, School of Natural Sciences, University of Zambia, Great East Road Campus, Box 32379, Lusaka, Zambia}
\affiliation{\abilene}
\affiliation{\augie}
\affiliation{\banaras}
\affiliation{\barc}
\affiliation{\baruch}
\affiliation{\bnlcoll}
\affiliation{\bnlphys}
\affiliation{\caucr}
\affiliation{\charlesczech}
\affiliation{\ciae}
\affiliation{\cns}
\affiliation{\colorado}
\affiliation{\columbia}
\affiliation{\czechtech}
\affiliation{\debrecen}
\affiliation{\elte}
\affiliation{\ewha}
\affiliation{\famu}
\affiliation{\fsu}
\affiliation{\gsu}
\affiliation{\hiroshima}
\affiliation{\howard}
\affiliation{\ihepprot}
\affiliation{\illuiuc}
\affiliation{\inrras}
\affiliation{\instpasczech}
\affiliation{\isu}
\affiliation{\jaea}
\affiliation{\jeonbuk}
\affiliation{\jyvaskyla}
\affiliation{\kek}
\affiliation{\korea}
\affiliation{\kurchatov}
\affiliation{\kyoto}
\affiliation{\lawllnl}
\affiliation{\losalamos}
\affiliation{\lund}
\affiliation{\lyon}
\affiliation{\maryland}
\affiliation{\mass}
\affiliation{\mate}
\affiliation{\michigan}
\affiliation{\miss}
\affiliation{\muhlenberg}
\affiliation{\nara}
\affiliation{\natmephi}
\affiliation{\newmex}
\affiliation{\nmsu}
\affiliation{\northcg}
\affiliation{\ohio}
\affiliation{\ornl}
\affiliation{\orsay}
\affiliation{\peking}
\affiliation{\pnpi}
\affiliation{\pusan}
\affiliation{\riken}
\affiliation{\rikjrbrc}
\affiliation{\rikkyo}
\affiliation{\saispbstu}
\affiliation{\seoulnat}
\affiliation{\stonybrkc}
\affiliation{\stonycrkp}
\affiliation{\tenn}
\affiliation{\texsu}
\affiliation{\titech}
\affiliation{\tsukuba}
\affiliation{\vandy}
\affiliation{\weizmann}
\affiliation{\wigner}
\affiliation{\yonsei}
\affiliation{\zagreb}
\affiliation{\zambia}
\author{U.~Acharya} \affiliation{\gsu} 
\author{A.~Adare} \affiliation{\colorado} 
\author{C.~Aidala} \affiliation{\michigan} 
\author{N.N.~Ajitanand} \altaffiliation{Deceased} \affiliation{\stonybrkc} 
\author{Y.~Akiba} \email[PHENIX Spokesperson: ]{akiba@rcf.rhic.bnl.gov} \affiliation{\riken} \affiliation{\rikjrbrc} 
\author{M.~Alfred} \affiliation{\howard} 
\author{V.~Andrieux} \affiliation{\michigan} 
\author{N.~Apadula} \affiliation{\isu} \affiliation{\stonycrkp} 
\author{H.~Asano} \affiliation{\kyoto} \affiliation{\riken} 
\author{B.~Azmoun} \affiliation{\bnlphys} 
% ...
\collaboration{PHENIX Collaboration}  \noaffiliation

\date{\today}


\begin{abstract}


Small nuclear collisions are mainly sensitive to cold-nuclear-matter 
effects; however, the collective behavior observed in these collisions 
shows a hint of hot-nuclear-matter effects. The identified-particle 
spectra, especially the $\phi$ mesons which contain strange and 
antistrange quarks and have a relatively small hadronic-interaction 
cross section, are a good tool to study these effects. The PHENIX 
experiment has measured $\phi$ mesons in a specific set of small 
collision systems $p$$+$Al, $p$$+$Au, and $^3$He$+$Au, as well as 
$d$$+$Au [Phys. Rev. C {\bf 83}, 024909 (2011)], at 
$\sqrt{s_{_{NN}}}=200$ GeV. The transverse-momentum spectra and 
nuclear-modification factors are presented and compared to 
theoretical-model predictions. The comparisons with different 
calculations suggest that quark-gluon plasma may be formed in these 
small collision systems at $\sqrt{s_{_{NN}}}=200$ GeV.  However, the 
volume and the lifetime of the produced medium may be insufficient for 
observing strangeness-enhancement and jet-quenching effects. The 
comparison with calculations suggests that the main production 
mechanisms of $\phi$ mesons at midrapidity may be different in $p$$+$Al 
versus $p/d/^3$He$+$Au collisions at $\sqrt{s_{_{NN}}}=200$ GeV. While thermal 
quark recombination seems to dominate in $p/d/^3$He$+$Au collisions, 
fragmentation seems to be the main production mechanism in $p$$+$Al collisions.

\end{abstract}

\maketitle

\section{INTRODUCTION}

Quantum chromodynamics (QCD) predicts the existence of a state of 
matter, called the quark gluon plasma (QGP) where quarks and gluons are 
unbounded, at either high temperature or high baryon density. 
Relativistic ion collisions provide unique opportunities to study 
properties and characteristics of the QGP in laboratory experiments, 
which is one of the main goals of the PHENIX experiment~\cite{QGP}. The 
experimental evidences of formation of QGP at \sqsn = 200 GeV have been 
observed in large collision systems such as Au$+$Au and 
Cu$+$Cu~\cite{PhidAuCuCuAuAu}, while the observables in \pp collisions 
are consistent with perturbative QCD (pQCD) calculations which describe 
primordial processes.  The specific set of small collision systems 
available at the Relativistic Heavy Ion Collider (RHIC) at \sqsn = 200 
GeV provides an opportunity to investigate the minimal conditions 
(temperature and/or baryon density) sufficient for QGP formation.

It is believed (e.g., see Ref.~\cite{QGP}) that in small collision 
systems (such as \pal, \pau, \dau, and \heau), where energy and/or 
baryon density are not high enough to form a QGP, multiparticle 
production in the final state may occur without a QCD phase transition. 
The cold-nuclear-matter effects~\cite{CNM, CNMdAu} seem to play a 
predominant role in small-system collisions at \sqsn = 200 GeV. These 
effects include multiple-parton scattering, nuclear absorption and 
modification of the initial parton-distribution functions (PDFs) in nuclei. 
However, recent studies on elliptic and triangular flow in small systems 
suggest that QGP could be produced in \pdheau 
collisions~\cite{FlowSmallSystems}. The flow measurements are 
well-explained by hydrodynamic model and are consistent with QGP droplet 
formation~\cite{HydroFlow}.

Additionally, the studies of \Jpsi~\cite{J_psi}, \PsiSS~\cite{Psi_SS}, 
and charged-hadron~\cite{Pseudorapidity_flow_small_systems, 
Charged_hadrons_small_syst} production at backward rapidity provide 
evidences of final-state effects not only in \pdheau, but also in \pal 
collisions. Nonetheless, these effects are weaker in \pal, than in 
\pdheau collisions.

The enhanced production of strange or hidden-strange hadrons in 
high-energy heavy-ion collisions, as compared to the appropriately 
scaled \pp collisions, is a direct consequence of the process of 
chemical equilibration of strange quarks in 
QGP~\cite{StrangenessEnhancenment}. Thus, measurement of hadrons 
containing (anti)strange quarks has been established as a promising 
method of detecting the QGP. Recently published ratios of strange to 
nonstrange hadron yields observed at the Large Hadron 
Collider~\cite{ALICE_Nature} show a smooth transition from 
elementary \pp collisions at the higher center-of-mass energy of \sqsn = 7 
TeV, via $p$$+$Pb collisions at \sqsn = 5.02 TeV, to heavy ion Pb$+$Pb 
collisions at lower energy \sqsn = 2.76 TeV, when studies as a function 
of the charged-particle multiplicity, $\langle dN_{ch}/d\eta \rangle$. 
This observation is interpreted as possible QGP formation in \pp or 
$p$$+$Pb collisions at high enough $\langle dN_{ch}/d\eta \rangle$ and 
demonstrates strangeness enhancement as a useful tool to study the onset 
of QGP formation. At RHIC, the strangeness enhancement in \dau 
collisions at \sqsn = 200 GeV is observed only in the Au-going direction 
($-2.2 < y < -1.2$) at $2<\pt<5\Gevc$, while in the $d$-going direction 
and at midrapidity, this effect is not observed within the 
uncertainties~\cite{Phi_dAu_FB, PhidAuCuCuAuAu}. Further measurements of 
strangeness enhancement in a broad set of small collision systems may 
provide an advantageous probe of QGP formation.

The strange-hadron yields also provide an additional degree of freedom, 
flavor, number of quarks, and mass, in the study of hadron production 
at high transverse momentum (\pt). The energy loss of hard-scattered 
partons in the QGP, called jet quenching, manifests itself as a suppression 
of hadron production at high \pt in relativistic ion collisions as 
compared to the expectations from elementary proton-proton 
collisions~\cite{JetQuenching}. The observation of both jet-quenching 
and strangeness-enhancement effects in various large systems (\auau and 
\cucu collisions at \sqsntwo~\cite{PhidAuCuCuAuAu}) suggests that QGP 
can be formed in such collisions. By now, in central collisions, the 
\vphi meson is less suppressed than other light-meson yields in the 
intermediate \pt range (2--5 GeV/$c$) whereas at higher \pt ($>$5 
GeV/$c$), all light mesons are suppressed in comparison to the \pp 
collisions and show similar suppression values~\cite{PhidAuCuCuAuAu}. 
Both strangeness enhancement and jet quenching observed in $A$$+$$A$ 
collisions are consistent with QGP formation, but are still  
under-explored in small collision systems at midrapidity and require 
further scrutiny.

The \vphi vector meson, which is the lightest nearly pure bound state of 
$s$ and $\bar{s}$ quarks~\cite{RevPartPhys} and measurable up to high 
\pt, is considered a good probe for the study of both jet-quenching and 
strangeness-enhancement effects in relativistic ion collisions. The 
interaction cross section of the \vphi meson with nonstrange hadrons has 
a small value~\cite{PhiQGP}. The data on coherent \vphi meson 
photo-production show that 
$\sigma_{\vphi N}{\approx}10~mb$~\cite{Review_phi,Phi_cross_section}. 
Additionally, because the lifetime (42 fm/c~\cite{PhiQGP}) is 
longer than the QGP ($\approx$5~fm/$c$~\cite{QGP}), \vphi mesons will 
decay mostly outside of the hot and dense matter and its daughters will 
not have much time to rescatter in the hadronic phase. Therefore, 
\vphi-meson production is expected to be less affected by the 
later-stage hadronic interactions in the evolution of the system formed 
in relativistic ion collisions.  Consequently, properties of the \vphi 
meson are primarily controlled by the conditions in the early partonic 
phase and, hence, can be considered a clean probe to investigate the 
properties of matter created in relativistic ion collisions. The \vphi 
meson has a mass of $\approx$1~GeV~\cite{RevPartPhys} which is 
comparable to the mass of the lightest baryons, such as protons.

This paper presents invariant \pt spectra and nuclear-modification 
factors of \vphi mesons in \pal, \pau, \dau, and \heau collisions at 
\sqsn = 200 GeV. The comparisons of obtained results to previous 
light-hadron-production measurements in small systems and to different 
model calculations are provided for better understanding of the 
underlying processes.

\section{DATA ANALYSIS}

A detailed description of the PHENIX experimental set-up can be found elsewhere ~\cite{PHENIXoverview}. The beam-beam counters (BBC)~\cite{BBC} 
are used for the centrality definition, the determination of collision 
vertex along the beam axis (the $z$-vertex), and the event start time. 
The BBCs cover the pseudorapidity range $3.0<|\eta|<3.9$.  The 
minimum-bias (MB) interaction trigger is also provided by the BBCs by 
requiring at least one inelastic nucleon-nucleon collision with the 
simultaneous detection of charged particles in both south BBC 
(Au[Al]-going direction) and north BBC ($p$[$^3$He]-going direction). 
The event vertex is required to be within $|z_{\text{vertex}}|<30$ cm of 
the nominal interaction region.

Two central arms (east and west) of the PHENIX detector are used for 
electron, photon, and charged-hadron measurements. They each cover 
$|\eta|<0.35$ in pseudorapidity and $90^o$ in azimuthal angle. The 
central arms include a particle-tracking system~\cite{TrackingSystem}, 
which comprises drift chambers and pad chambers.

Charged particle identification (PID) is performed by simultaneous 
measurement of momentum, flight time, and path-length. The flight time 
is measured by the time-of-flight detector in the east part of the 
central arm spectrometer (TOF-E)~\cite{ToF, ToF2}.

The data sets used in the analysis are collected from \pal and \pau 
collisions by the PHENIX detector in 2015 and \heau collisions collected 
in 2014 at center-of-mass energy \sqsntwo. The integrated luminosities 
of the data sets used in this analysis are $1.27~pb^{-1}$ in \pau, 
$3.87~pb^{-1}$ in \pal, and $134~ nb^{-1}$ in \heau collisions.

Centrality selection is performed with the BBCs using the Glauber Monte 
Carlo procedure described in~\cite{CentralitySelection}, wherein the 
charge in the BBCs is assumed to be proportional to the number of 
participating nucleons \Npart. In the current study, the distributions 
measured in the south BBC are used, which is the direction of the larger 
nucleus (Al or Au). The BBC charge is assumed to follow a 
negative-binomial distribution (NBD) with a mean of \Npart and the 
remaining NBD parameters determined from a $\chi ^2$ minimization of the 
combined Glauber+NBD calculation with respect to the data. The BBC 
distributions are divided into equal probability bins, and the 
corresponding Glauber distributions are used to calculate \Npart as well 
as the number of binary nucleon-nucleon collisions \Ncoll , which are 
shown in Table~\ref{table:MC_parameters}.

       \begin{table}[tbh]
 \caption{\label{table:MC_parameters}
   Summary of the \Ncoll ,  \Npart , and $f_{\rm bias}$ values calculated using Glauber Monte Carlo simulation.}
 \begin{ruledtabular} \begin{tabular}{ccccc}
Collision & Centrality & \Ncoll    &  \Npart    & $f_{{\rm bias}}$       \\ \hline
\pal & 0\%--72\%     & 2.1$\pm$0.1  & 3.1$\pm$0.1    & 0.80$\pm$0.02    \\
& 0\%--20\%     & 3.4$\pm$0.3  & 4.4$\pm$0.3    & 0.81$\pm$0.01    \\
& 20\%--40\%    & 2.3$\pm$0.1  & 3.3$\pm$0.1    & 0.90$\pm$0.02    \\
& 40\%--72\%    & 1.6$\pm$0.1  & 2.6$\pm$0.1    & 1.05$\pm$0.04  \\
\\
\pau & 0\%--84\%     & 4.7$\pm$0.3  & 5.7$\pm$0.3    & 0.86$\pm$0.01 \\
& 0\%--20\%     & 8.2$\pm$0.5  & 9.2$\pm$0.5    & 0.90$\pm$0.01   \\
& 20\%--40\%    & 6.1$\pm$0.4  & 7.1$\pm$0.4    & 0.98$\pm$0.01   \\
& 40\%--84\%    & 3.4$\pm$0.2  & 4.4$\pm$0.2    & 1.01$\pm$0.04 \\
\\
\heau & 0\%--88\%     & 10.4$\pm$0.7 & 11.4$\pm$0.5 & 0.89$\pm$0.01   \\
& 0\%--20\%     & 22.3$\pm$1.7 & 21.1$\pm$1.3 & 0.95$\pm$0.01   \\
& 20\%--40\%    & 14.8$\pm$1.1 & 15.4$\pm$0.9 & 1.01$\pm$0.01   \\
& 40\%--60\%    & 8.4$\pm$0.6  & 9.5$\pm$0.6  & 1.02$\pm$0.01   \\
& 60\%--88\%    & 3.4$\pm$0.3  & 4.6$\pm$0.3  & 1.03$\pm$0.05   \\
 \end{tabular} \end{ruledtabular}
 \end{table}

The determination of hadron yields in centrality bins has a known bias 
effect (see Ref.~\cite{Centrality}). This effect results from the diffractive 
portion of the \pp collision, constituent \pal or \pdheau collision, and 
manifests itself as a bias towards nondiffractive collisions, where 
higher charge is deposited in the BBC, and hence towards larger 
centrality. Increased trigger efficiency is correlated with a 1.55 times 
larger BBC multiplicity~\cite{Pseudorapidity_flow_small_systems}. Bias 
effects were removed via correction factors $f_{{\rm bias}}$ that are 
calculated using a Glauber+NBD approach and following the detailed 
procedure described in Ref.~\cite{Centrality}.

The \vphi-meson-production measurement is conducted via the kaon 
($K$-meson) decay channel.  The values of \vphi meson mass, width 
($\Gamma$) and branching ratio (Br) of 
$\vphi \rightarrow K^+ K^-$ decay can be found in~\cite{RevPartPhys}.

\begin{figure}[!ht]
    \includegraphics[width=0.96\linewidth]{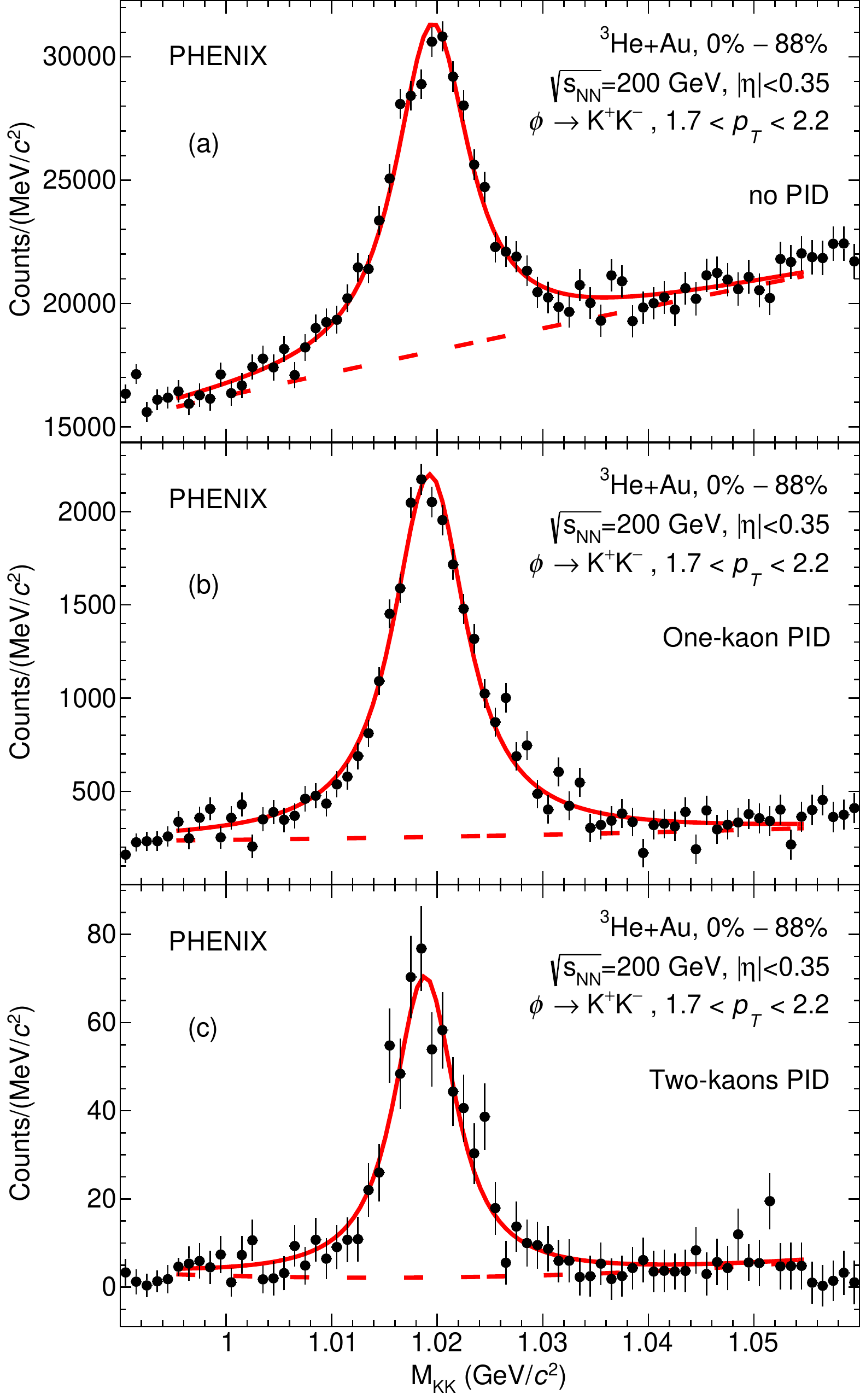}
\caption{
Examples of invariant-mass distributions for the $K^+K^-$ pairs in \heau 
collisions at \sqsn = 200 GeV, obtained with the (a) no PID, (b) 
one-kaon PID, and (c) two-kaons PID methods after subtraction of the 
combinatorial background estimated using the event-mixing technique. 
Plots correspond to integrated \pt for $1.7<\pt<2.2$ \Gevc. Spectra are 
fitted to the sum of a Breit-Wigner function convolved with a Gaussian 
function to account for the \vphi signal, and a polynomial function to 
account for the residual background.}
\label{fig:Minv}
    \end{figure}

    
Each charged track is paired with its opposite sign to reconstruct the 
invariant-mass spectrum in every selected centrality class and \pt bin. 
For every track, the three momentum components are determined with the 
help of the drift chambers and the first layer of the pad chambers. 
Then, the invariant mass and transverse momentum are calculated from the 
kinematics of two-particle decay. This, so called ``no~PID," technique 
is used for all collisions for \pt $>$ 2.2 \Gevc. To increase the 
signal to background ratio for \pt $<$ 2.2 \Gevc, one of the tracks is 
required to be identified as a kaon. The requirements of the charged 
track to be a kaon is determined by the TOF-E detector.  This so-called 
``one-kaon~PID" is used for \pt $<$ 2.2 \Gevc in \pau and \heau 
collisions. Additionally, to provide a cross check of the results and 
for estimation of systematic uncertainties for \heau collisions, both 
kaons were required to be identified (``two-kaons~PID"). For \pal 
collisions only the ``no~PID" technique is applied due to low 
statistics.  Figure~\ref{fig:Minv} shows examples of mass spectra 
obtained in \heau collisions using the three methods.

Invariant-mass spectra for opposite sign pairs comprise the \vphi-meson 
signal and the combinatorial background. The combinatorial background 
comprises correlated and uncorrelated parts. The event-mixing 
technique~\cite{EventMixing} is applied in order to subtract the 
uncorrelated background. After subtraction, the background 
invariant-mass distribution is fitted with a Gaussian function 
convoluted with Breit-Wigner function to describe the signal and a second order
polynomial function to describe the remaining correlated background from 
other particle decays ($K_{s}^{0}\rightarrow \pi^{+}\pi^{-}, ~\Lambda 
\rightarrow p\pi^{-},~ \rho \rightarrow \pi^{+}\pi^{-}, ~\omega 
\rightarrow \pi^{0}\pi^{+}\pi^{-}$ etc.). Gaussian $\sigma$ value, 
corresponding to mass resolution, is constrained to the
$\sigma_{exp}$ value derived using a full 
{\sc geant}~\cite{GEANT} simulation of the PHENIX detector with zero natural 
width of \vphi meson. The $\Gamma$ parameter of the Breit-Wigner function
is left as a free parameter in the fit of the simulated data, and its
extracted value $\Gamma_0$ is then used in the real data fitting to
constrain the $\Gamma$ parameter to fall within $\pm10\%$ of the 
$\Gamma_0$ value. The reconstruction efficiency 
($\varepsilon_{\rm rec}$) of the \vphi meson is determined using simulation with a 
\vphi meson PDG width $\Gamma$. The raw yields of \vphi mesons are 
obtained by integrating the invariant mass distribution in the range 
$\pm~9$ \Mevcc around the \vphi meson mass after combinatorial background 
subtraction as shown in Fig.~\ref{fig:Minv}.
  
\begin{table}[!htb]
\begin{minipage}{1.0\linewidth}
\caption{\label{table:Syst_pal}
Type B systematic uncertainties on the \vphi meson
invariant yields in \pal collisions at \sqsn = 200~GeV}
\begin{ruledtabular} \begin{tabular}{rccc}
                \pt[GeV/$c$] & 1.45 & 3.45  & 3.95  \\
\hline
 Raw-yield extraction       & 18.1\%  & 9.9\%  & 11.2\%  \\
 Acceptance                  & 4.0\%  & 4.0\%  & 4.0\%   \\
 Reconstruction efficiency     & 3.0\%  & 3.0\%  & 3.0\% \\
 Momentum scale              & 0.6\%  & 3.0\%  & 3.6\%   \\
 Branching ratio             & 1.0\%  & 1.0\%  & 1.0\%   \\
\\
 Total type B              & 18.8\% & 11.5\% & 12.7\%  \\
 \end{tabular} \end{ruledtabular}
 \end{minipage}
\begin{minipage}{1.0\linewidth}
 \caption{
Type B systematic uncertainties on the \vphi meson
invariant yields in \pau collisions at \sqsn = 200 GeV}
\label{table:Syst_pau}
 \begin{ruledtabular} \begin{tabular}{rccc}
               \pt[GeV/$c$] & 1.1 & 1.95 & 3.95       \\
\hline
 Raw-yield extraction      & 9.6\%  & 8.8\%   & 11.2\%   \\
Acceptance                 & 5.0\%  & 5.0\%   & 5.0\%    \\
Reconstruction efficiency  & 4.0\%  & 4.0\%   & 4.0\%    \\
Momentum scale             & 0.5\%  & 1.1\%   & 3.6\%    \\
Branching ratio            & 1.0\%  & 1.0\%   & 1.0\%    \\
\\
Total type B             & 11.6\%  &11.0\%  & 13.4\% \\
 \end{tabular} \end{ruledtabular}
\end{minipage}
\begin{minipage}{1.0\linewidth}
 \caption{\label{table:Syst_heau}
Type B systematic uncertainties on the \vphi meson
invariant yields in \heau collisions at \sqsn = 200 GeV}
 \begin{ruledtabular} \begin{tabular}{rcccc}
                \pt[GeV/$c$] & 1.1  & 1.95 & 5.5  & 7.0  \\
\hline
Raw-yield extraction      & 7.6\%  & 6.8\%  & 15.4\% & 14.8\% \\
Acceptance                & 4.0\%  & 4.0\%  & 4.0\%  & 4.0\%  \\
Reconstruction efficiency & 3.0\%  & 3.0\%  & 3.0\%  & 3.0\%  \\
Momentum scale            & 0.5\%  & 1.1\%  & 4.7\%  & 5.0\%  \\
Branching ratio           & 1.0\%  & 1.0\%  & 1.0\%  & 1.0\%  \\
\\
Total type B            & 9.2\%  & 8.6\%  & 16.9\% & 16.4\% \\
 \end{tabular} \end{ruledtabular}
\end{minipage}
\begin{minipage}{1.0\linewidth}
\caption{\label{table:typeC}
Type C systematic uncertainties on the \vphi meson
invariant yields in \pal, \pau, and \heau collisions at \sqsn = 200~GeV}
\begin{ruledtabular} \begin{tabular}{rcccc}
\pal & Centrality    & 0\%--20\% & 40\%--72\% & 0\%--72\% \\ 
     & Total type C  & 10.5\%   &  9.2\%   &  7.7\%    \\
\\
\pau & Centrality    & 0\%--20\% & 40\%--84\% & 0\%--84\%  \\ 
     & Total type C  & 6.6\%  & 7.4\%  & 6.9\%    \\
\\
\heau & Centrality   & 0\%--20\% & 60\%--88\% & 0\%--88\%   \\ 
      & Total type C & 7.7\%    &  10.1\%    &  6.9\%    \\
\end{tabular} \end{ruledtabular}
\end{minipage}
\end{table}

The invariant spectra of \vphi meson in each transverse-momentum 
interval is calculated as
\begin{align}
	 \frac{1}{2\pi N_{\rm event}}\frac{d^2N}{\pt d\pt dy} &=\frac{f_{\rm bias}}{2\pi\pt} \nonumber \\
 &\times \frac{N_{\rm raw}}{N_{\rm event} {\rm Br}\cdot \varepsilon _{\rm rec}(\pt) \Delta \pt \Delta y},
   \end{align}
where $N_{{\rm raw}}$ is the number of \vphi mesons detected by the 
experimental setup (raw yield), $N_{{\rm event}}$ is the number of 
analyzed events, Br is the branching ratio of $\vphi \rightarrow K^+ 
K^-$ decay, and $\varepsilon _{\rm rec}(\pt)$ corrects for the 
limited acceptance of the detector and the \vphi meson reconstruction 
efficiency.

Nuclear-modification factors are calculated as 
 \begin{equation}\label{RxA}
	 R_{xA}=\frac{ \sigma^{\rm inel}_{pp}}{\Ncoll}\cdot \frac{d^2N_{xA}/dydp_T}{d^2\sigma_{pp}/dydp_T},
 \end{equation}
where $d^2N_{xA}/dydp_T$ is the per-event yield of particle production 
in $x+A$ collisions, $d^2\sigma_{pp}/dydp_T$ is the production cross 
section in \pp collisions, and $\sigma^{\rm inel}_{pp}=42.2$ mb~\cite{pp} is 
the total inelastic proton-proton cross section. The \pp reference data 
used in the analysis is taken from~\cite{pp}.

There are three types of systematic uncertainties: type A 
(point-to-point uncorrelated); type B (point-to-point correlated), which 
can change the shape of the spectrum in a smooth way as a function of 
\pt and type C (global or normalization), which can only move all data 
points up or down by the same amount. The uncertainties of type A are 
dominated by the statistical precision of the data. Uncertainty of type 
B includes acceptance, reconstruction efficiency and momentum scale 
uncertainties, and uncertainty in the raw-yield extraction, which are 
evaluated by varying the identification approaches, fit parameters and 
the parameterization of the residual background. The trend in raw yield extraction uncertainty values at low \pt is mostly driven by the increasing signal to background ratio with increasing \pt, and at high \pt by worsening detector mass resolution  and lower statistics. The various 
normalization correction terms have type C uncertainties. Uncertainty of 
type C includes \Ncoll, $f_{{\rm bias}}$ uncertainties, uncertainty 
caused by event overlap (0.9$\%$ for \heau, 2.2$\%$ for \pau, and 
5.5$\%$ for \pal), which might arise during the same bunch crossing, and 
uncertainty in normalization for the \pp cross section equal to $\approx 
9.7\%$. The uncertainties are examined in each centrality class for 
\pal, \pau, and \heau collisions and are found to be consistent among 
all centrality classes.

Tables~\ref{table:Syst_pal},~\ref{table:Syst_pau}, 
and~\ref{table:Syst_heau} present typical values of the estimated type-B 
systematic uncertainties and Table~\ref{table:typeC} shows those for 
type C.  In all three systems, the total systematic error is dominated 
by raw-yield extraction uncertainty.

\section{RESULTS}

\begin{figure*}[!ht]
\includegraphics[width=0.99\linewidth]{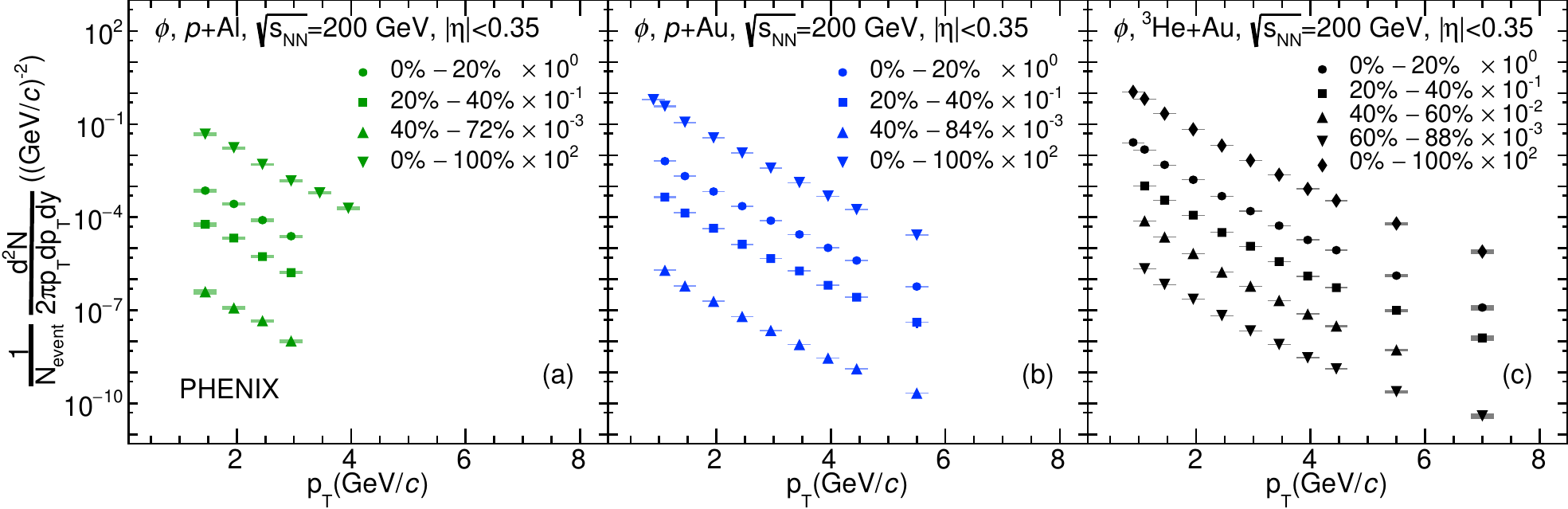}
\caption{
Invariant transverse momentum spectra measured for \vphi mesons in (a) \pal, 
(b) \pau and (c) \heau collisions at \sqsntwo at midrapidity. The 
statistical uncertainties are shown by vertical bars, which are smaller than 
the size of the symbols, and the systematic uncertainties are indicated by 
rectangles, which are depicted wide to make them visible.
}
\label{fig:Spectra}
\end{figure*}

Figure~\ref{fig:Spectra} shows the invariant transverse momentum spectra 
of \vphi mesons in \pal, \pau and \heau collisions at \sqsntwo at 
midrapidity $|\eta|<0.35$, in four centrality bins in \pal and \pau and 
for five centrality bins in \heau collisions.


\begin{figure*}[!ht]
\includegraphics[width=0.96\linewidth]{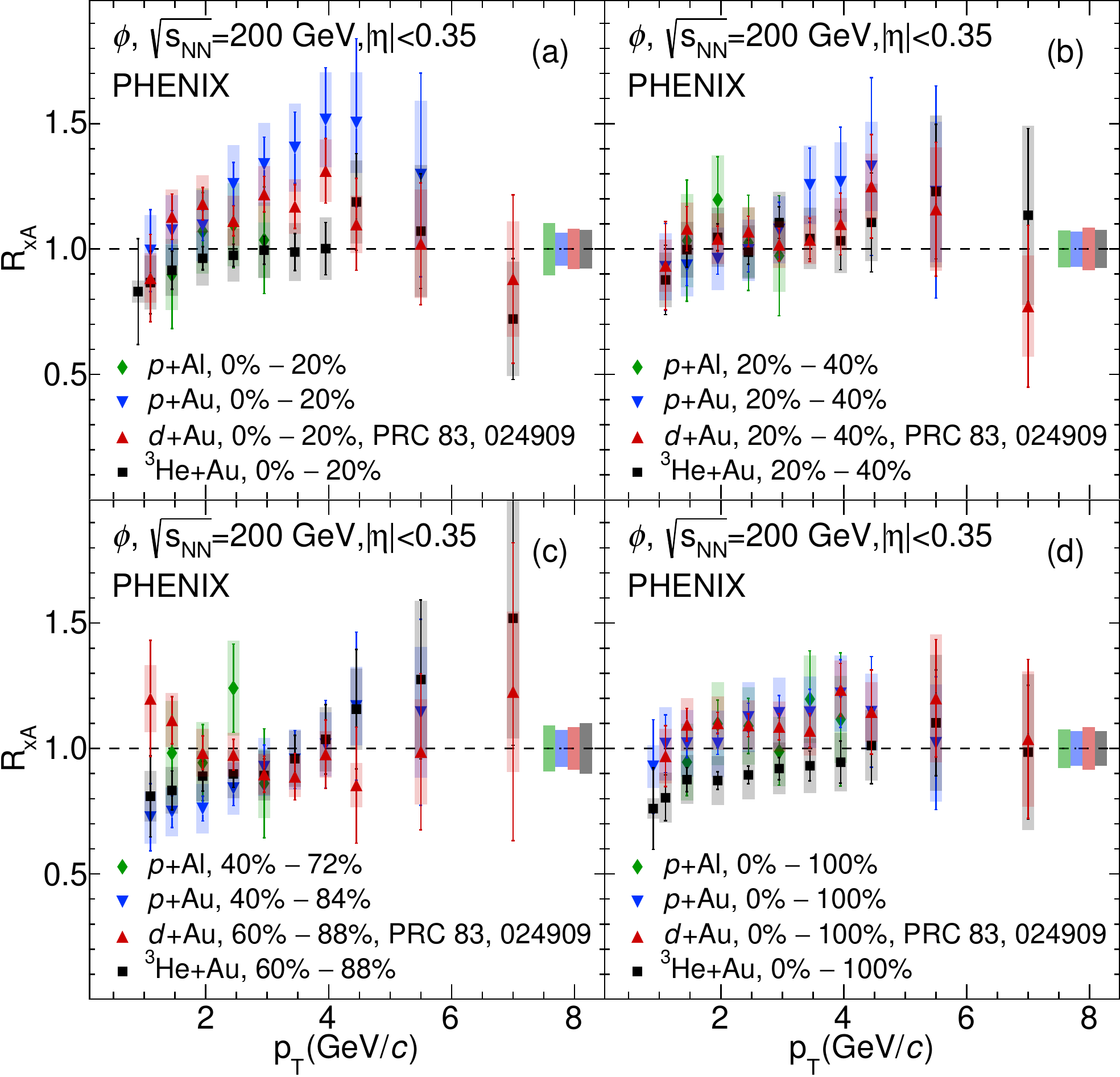}
\caption{
Comparison of \vphi-meson nuclear-modification factors in \pal, \pau, 
\dau~\protect\cite{PhidAuCuCuAuAu}, and \heau collisions at \sqsntwo at 
midrapidity. Here and in following figures, the statistical uncertainties 
are shown by vertical bars and the systematic uncertainties are indicated by 
rectangles, which are depicted wide to make them visible.  The normalization 
uncertainty from \pp of about 9.7\% is not shown~\cite{pp}.
}
\label{fig:RxA}
\end{figure*}

Figure~\ref{fig:RxA} shows \vphi meson nuclear-modification factors \rxa 
measured in \pal, \pau, \dau and \heau collisions at \sqsn = 200 GeV at 
midrapidity. The normalization uncertainty from \pp ($\approx9.7\%$) is not 
shown~\cite{pp}. From comparing \pal, \pau, \dau and \heau results, an 
ordering of \vphi-meson \rxa might be seen in the intermediate \pt range in 
the most-central (0\%--20\%) and MB (0\%--100\%) collisions: 
$R_{^{3}{\rm HeAu}}<R_{{\rm dAu}}<R_{{\rm pAu}}$. Also at high \pt, a 
hint of suppression in central collisions, and a hint of enhancement in 
peripheral collisions is observed.  Nonetheless, the \vphi-meson \rxa are 
equal to unity within large uncertainties. Similar results were 
previously obtained for \pio production in small collision systems and 
was explained by conservation of energy~\cite{PPG202}. The production of 
high-energy particles (with a large transverse momentum), by virtue of 
the conservation of energy, leads to a decrease in multiplicity in the 
collision~\cite{BiasCentrality} and hence~\cite{Centrality} possibly 
incorrectly categorizing some central collisions as peripheral 
collisions.  This effect might cause, at high \pt, a hint of \rxa 
suppression in central collisions and a hint of \rxa enhancement in 
peripheral collisions. This suggests that conclusions about energy loss 
in small collision systems cannot be drawn due to insufficient 
experimental precision and further careful theoretical treatment is 
required.


\begin{figure*}[!ht]
\includegraphics[width=0.99\linewidth]{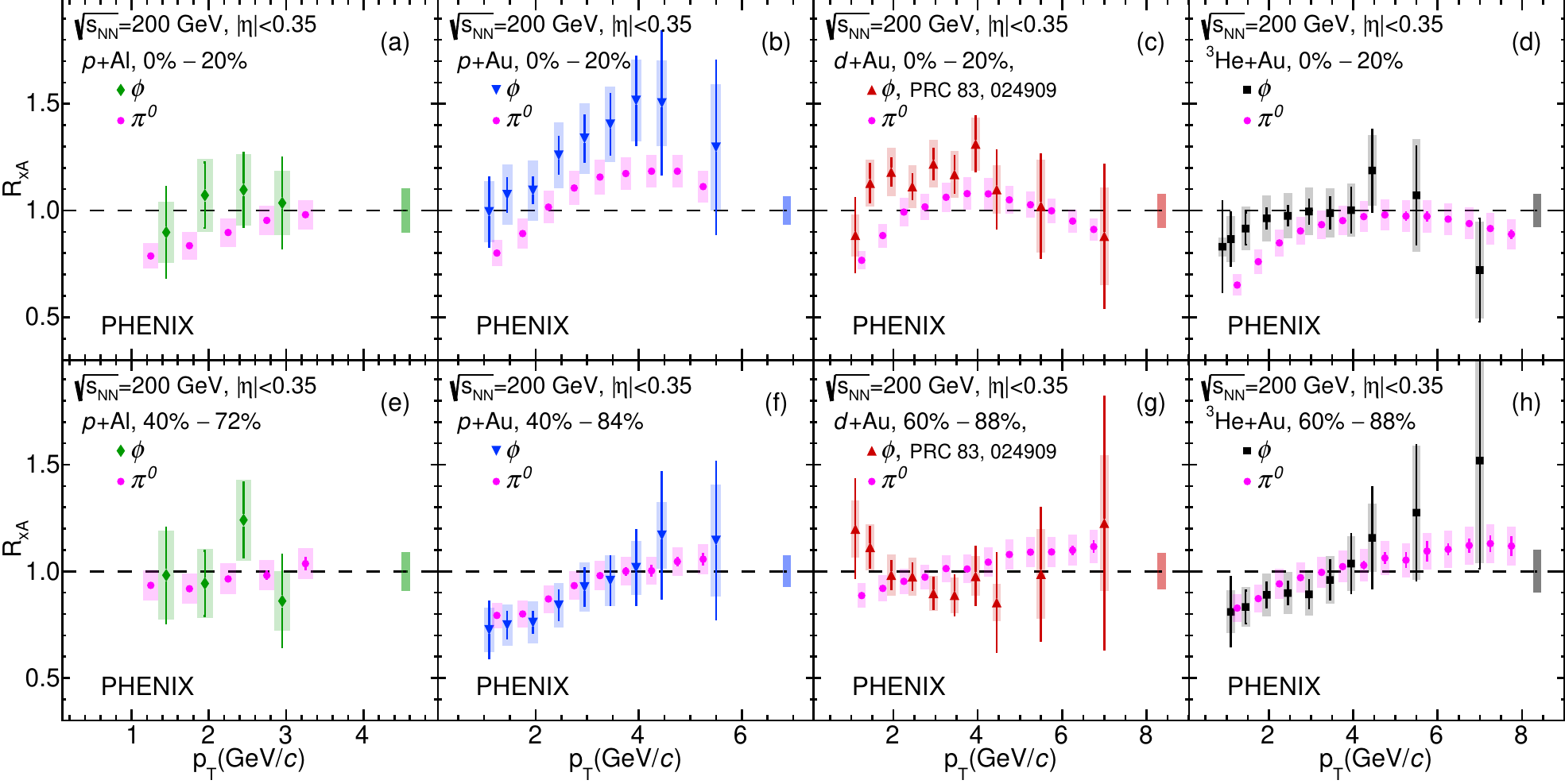}
\caption{The comparison of \vphi meson to \pio-meson (from 
Ref.~\cite{PPG202}) nuclear-modification factors in \pal, \pau, 
\dau~\protect\cite{PhidAuCuCuAuAu}, and \heau (a) to (d) central and (e) 
to (h) peripheral collisions at \sqsn = 200 GeV at 
midrapidity. The normalization uncertainty from \pp ($\approx9.7\%$) is 
not shown. }
\label{fig:Pi}
\end{figure*}

Figure~\ref{fig:Pi} shows the comparison of \vphi meson and \pio-meson 
nuclear-modification factors~\cite{PPG202} measured in \pal, \pau, 
\dau~\cite{PhidAuCuCuAuAu}, and \heau collisions at \sqsntwo at midrapidity.  
Because \vphi meson contains $s$ and $\bar{s}$ quarks and \pio comprises of 
$u$ and $d$ quarks, this comparison can reveal the possible strangeness 
enhancement effect. Panels (a) to (d) show the results for the most central 
collisions and and panels (e) to (h) show the results for the most 
peripheral collisions. In the \vphi meson \pt-range up to 8.0 GeV/$c$, \vphi 
and \pio-mesons nuclear-modification factors are in agreement within their 
uncertainties for the different collision systems.  The \vphi meson 
production in the most central collisions shows a trend to less suppression 
or larger enhancement than the \pio meson production at moderate \pt, 
however it cannot be concluded due to large systematic uncertainties. In 
heavy ion collisions (\auau and \cucu), in the most central collisions, the 
\vphi meson \rxa shows less suppression than \pio-meson in the intermediate 
\pt range of $2 < \pt (\Gevc)< 5$~\cite{PhidAuCuCuAuAu}. This result is 
qualitatively consistent with quark coalescence from QGP 
models~\cite{Recombination1,Recombination2}. The observation of 
strangeness enhancement in small collision systems at midrapidity 
cannot be concluded due to large systematic uncertainties.

\begin{figure*}[!ht]
\includegraphics[width=0.6\linewidth]{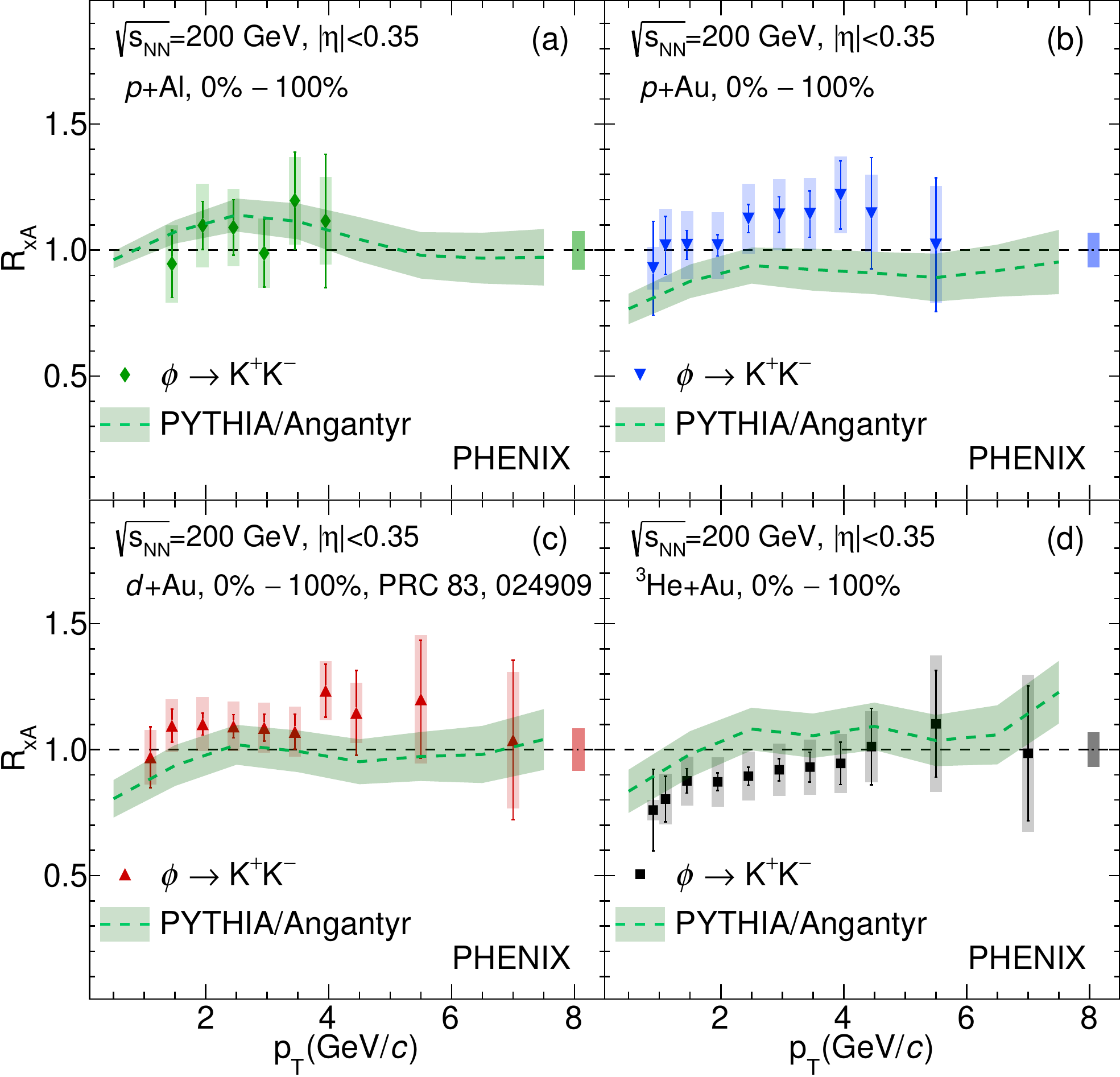}
\caption{\label{fig:PYTHIA}
Experimental results on \vphi meson production in (a) \pal, (b) \pau, 
(c) \dau~\protect\cite{PhidAuCuCuAuAu}, and (d) \heau collisions at 
\sqsn = 200 GeV at midrapidity ($|\et|<0.35$) and comparisons to 
{\sc pythia}/Angantyr~\protect\cite{PYTHIA_ANGANTYR} model predictions.}
\end{figure*}

To separate collective and noncollective phenomena and to study the 
\vphi meson production mechanism, the data are compared to calculations 
using {\sc pythia}/Angantyr~\cite{PYTHIA_ANGANTYR}, 
Eskola-Paakkinen-Paukkunen-Salgado ({\sc epps16})~\cite{EPPS16}, 
coordinated-theoretical-experimental project on QCD ({\sc ncteq15}) nuclear
PDF~\cite{nCTEQ15}, and a multiphase transport ({\sc ampt})~\cite{AMPT} 
models for both default [def] and string-melting [sm].

{\sc pythia}8.303~\cite{PYTHIA8} was developed based on leading-order pQCD 
calculations with soft-hadron production matching the observed data from 
\pp collisions at different energies. To further develop its framework, 
Angantyr was created to include heavy ion collisions in the same {\sc pythia} 
framework without introducing a new state of matter (collective 
behavior).

The first step is to establish the inclusive hadron spectrum in \pp 
collisions at \sqs = 200 GeV from the {\sc pythia}v8.303. Then, the \vphi 
meson spectra were estimated from the {\sc pythia}/Angantyr in \pal, \pau, 
\dau or \heau collision at the same collision energy. The parameters 
used in the event generation of {\sc pythia} are listed in 
Table~\ref{table:PYTHIA_parameters}. The multiplication factor for 
multiparton interactions is introduced to match $\eta$-dependent 
multiplicity distribution in \pp collisions at \sqsn = 200 GeV in {\sc pythia} 
calculations and experimental data~\cite{dNch_deta}. The \rxa were then 
calculated with the \Ncoll values taken from {\sc pythia}/Angantyr, which are 
listed in Table~\ref{table:PYTHIA_Ncoll}.

 \begin{table}[tbh]
     \caption{\label{table:PYTHIA_Ncoll}
\Ncoll values obtained from {\sc pythia}}8.303~\protect\cite{PYTHIA8}.
     \begin{ruledtabular} \begin{tabular}{cccc}
    \pal & \pau & \dau & \heau \\\hline
     2.1 &  4.2 &  6.2 & 7.9   \\
     \end{tabular} \end{ruledtabular}
     \end{table}

Systematic uncertainties for {\sc pythia}/Angantyr calculations include 
the uncertainty of the PDFs variation and uncertainty in total $x$+A 
cross section. Figure~\ref{fig:PYTHIA} shows the comparison of 
experimental results on \vphi meson production in \pal, \pau, \dau and 
\heau at \sqsn = 200 GeV to {\sc pythia}/Angantyr model predictions. The 
results shown for the MB collisions suggest that {\sc pythia}/Angantyr 
calculations describe the experimental results within uncertainties, 
however predict the reverse \rxa ordering: 
$R_{{\rm pAu}}<R_{{\rm dAu}}<R_{^{3}{\rm HeAu}}$.  Despite the agreement 
of \rxa experimental values with {\sc pythia}/Angantyr calculations,
the same calculations have discrepancies 
with experimental results on the \vphi-meson invariant-\pt spectra in 
\pp~\cite{Phi_pp_Pythia} and all considered systems at \sqsntwo 
at midrapidity.
  
\begin{figure*}[!ht]
\includegraphics[width=0.6\linewidth]{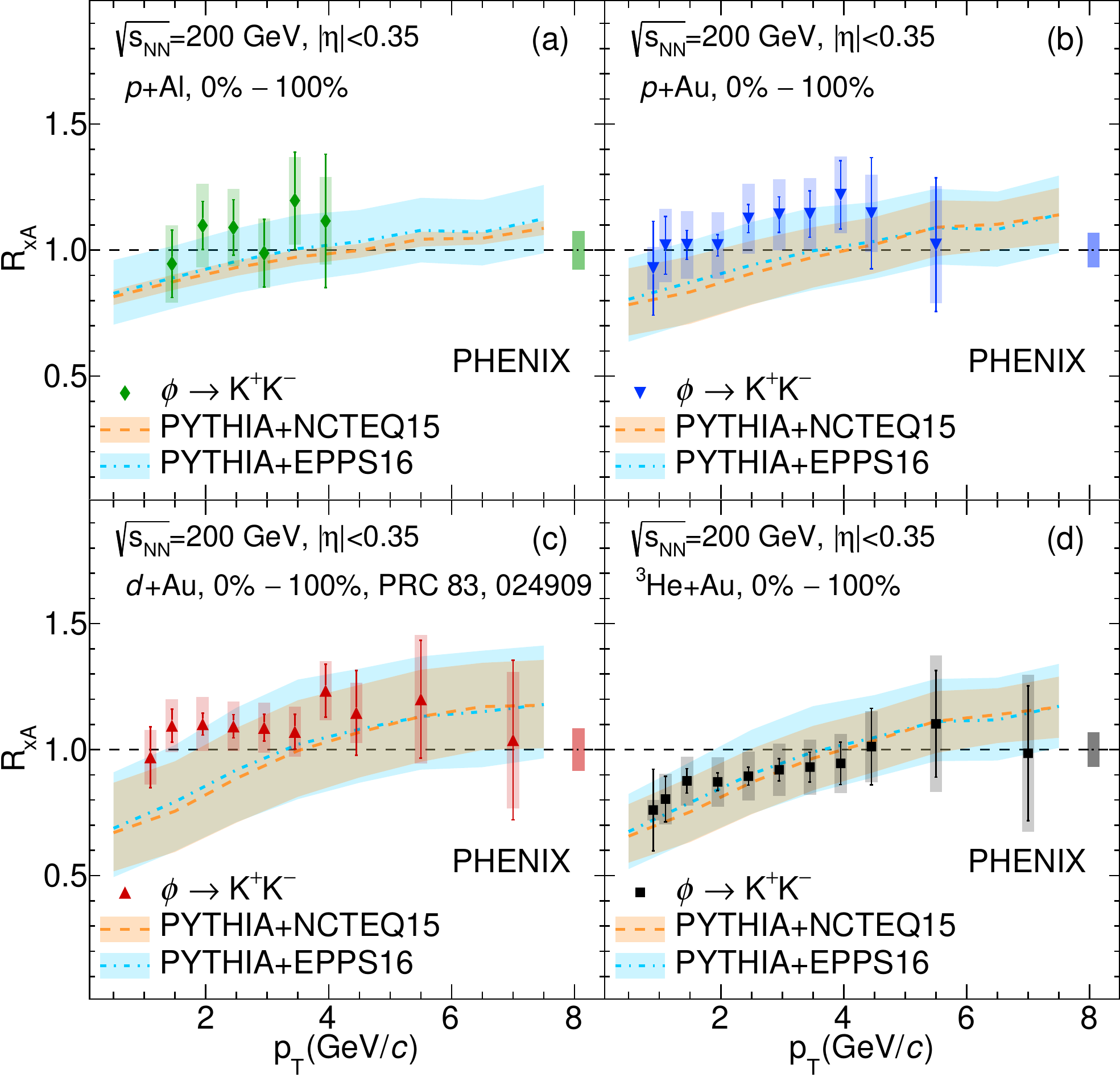}
\caption{\label{fig:nPDF}
Experimental results on \vphi-meson production in 
(a) \pal, (b) \pau, (c) \dau~\protect\cite{PhidAuCuCuAuAu}, and (d) 
\heau collisions at \sqsntwo at midrapidity ($|\et|<0.35$) and 
comparisons to {\sc epps16}~\protect\cite{EPPS16} and 
{\sc ncteq15}~\protect\cite{nCTEQ15} nuclear PDF calculations.}
\end{figure*}

Figure~\ref{fig:nPDF} shows the experimental data compared to 
calculations based on {\sc ncteq15} nPDF~\cite{nCTEQ15} 
and {\sc epps16} nPDF~\cite{EPPS16} interfaced with {\sc pythia}8.303. 
Both {\sc ncteq15}, and {\sc epps16} nPDF results show conformity with 
experimental data within uncertainties.  However, the nPDF calculations fail 
to predict the experimental ordering of \vphi meson \rxa at moderate 
\pt, as was previously observed for \pio production. The different 
trends of the nPDF calculations compared to the experimental data 
suggest that the nuclear modification in \pdheau collisions might 
involve some mechanism(s) additional to nPDF.

 \begin{table*}[tbh]
 \caption{\label{table:PYTHIA_parameters}
   Parameters used in {\sc pythia} }
 \begin{ruledtabular}
 \begin{tabular}{ccc}
parameter  & value         &  description                                           \\ \hline
SoftQCD:   & all = on        & All soft QCD processes                                 \\
           &               & Used for {\sc pythia} /Angantyr calculations                  \\
\\
           & inelastic = on  & All soft QCD processes, except for elastic             \\
 &   & Used for {\sc ncteq15+pythia} and {\sc epps16+pythia} calculations \\
\\
PDF:pSet   & 8             & {\sc cteq6l1} parton-distribution function                   \\ 
\\
MultipartonInteractions:Kfactor  & 0.5 & Multiplication factor for multiparton interaction \\ 
 \end{tabular}
 \end{ruledtabular}
 \end{table*}

\begin{figure*}[!ht]
\includegraphics[width=0.6\linewidth]{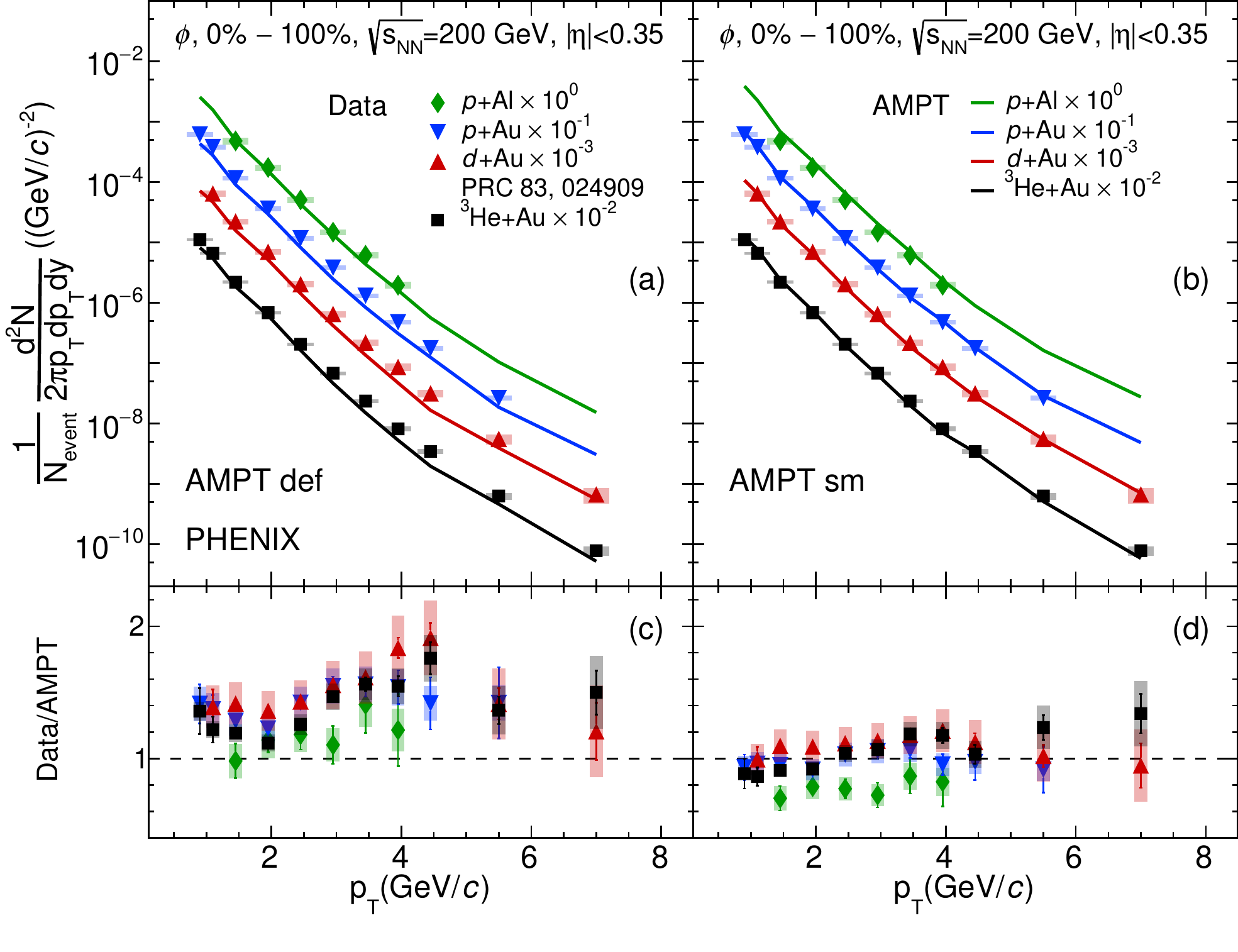}
\caption{\label{fig:AMPT}
Experimental results on \vphi meson invariant \pt spectra in \pal, \pau, 
\dau, and \heau collisions at \sqsn = 200 GeV at midrapidity 
($|\et|<0.35$) and comparisons to (a) default [def] and (b) string 
melting [sm] versions of the {\sc ampt}-model predictions. Panels (c) and (d) 
show data to {\sc ampt}-calculation ratios; the markers, error bars, and 
error boxes are the same as for panels (a) and (b).}
\end{figure*}

The {\sc ampt} model~\cite{AMPT} includes the initial-partonic and 
final-hadronic matter, as well as the transition between the two phases.  
This model provides an opportunity to study the 
hadronization mechanism in relativistic ion collisions.
In the {\sc ampt}-default model, only minijet partons from processes 
evaluated by the pQCD are involved in the Zhang's parton 
cascade~\cite{ZhangPartonCascade} and are recombined with their parent 
strings when they stop interacting.  The resulting strings are converted 
to hadrons using the Lund string fragmentation model.

In the extended-string-melting version of the {\sc ampt} model, the 
strings, formed in the nonperturbative processes, melt into partonic 
degrees of freedom and a quark-coalescence model~\cite{AMPT} is used to 
combine partons into hadrons. The {\sc ampt} results were obtained using 
a parton-scattering cross section of 3.0 mb and incorporating the nuclear 
shadowing effect~\cite{AMPT}.

Figure~\ref{fig:AMPT} shows the comparison of experimental \vphi meson 
invariant \pt spectra in MB \pal, \pau, \dau, and \heau collisions to 
the predictions of default and string-melting {\sc ampt} calculations. 
The \pdheau results are well described in the frame of the 
string-melting version of the {\sc ampt} model. The ratios of 
\vphi-meson yields, measured in the experiment to {\sc ampt} calculations, are 
consistent with each other in \pdheau collisions and therefore, the 
{\sc ampt} model is able to predict the experimental ordering of \vphi 
meson \rxa. The default version calculations underpredict the 
experimental data for \pdheau collisions. In contrast, the string-melting 
version of the {\sc ampt}-model calculations seems to overpredict the 
\vphi-meson invariant spectra in \pal results, whereas the 
default-version calculations demonstrate more conformity. Therefore, the 
coalescence mechanism apparently plays a considerable role in 
hadronization in \pdheau collisions at \sqsntwo. This confirms previous 
studies of light-hadron production at RHIC, where some of QGP effects 
such as baryon enhancement~\cite{BaryonEnhancement} and reversed mass 
ordering of $v_{2}$ in small collision 
systems~\cite{AzimuthalAnisotropySmallSystems} have been interpreted in 
terms of recombination model of hadronization. However, the comparison 
of experimental data to theoretical model predictions in the current 
study suggests that in \pal at midrapidity the contribution of the 
coalescence mechanism in \vphi meson production is less significant.

\section{SUMMARY}

In summary, PHENIX has measured \vphi meson invariant transverse 
momentum spectra in $|\eta|<0.35$ in \pal, \pau and \heau collisions at 
\sqsntwo in the range $1.0 <$ \pt $< 4.2(6.25,7.75)$ GeV/$c$ for 
different centrality classes via the kaon decay channel. The 
nuclear-modification factors in these collision systems were also 
presented. These first measurements of \vphi meson production and its 
nuclear modification in highly asymmetric small collision systems at 
RHIC fill the gaps in \vphi meson measurements between previous results 
in \pp, \dau, and heavy-ion collisions.

In the most central and MB collisions in the intermediate \pt 
range \vphi meson nuclear-modification factors show a hint of ordering: 
$R_{^{3}{\rm HeAu}}<R_{{\rm dAu}}<R_{\rm pAu}$. In other 
centralities, \vphi meson \rxa exhibit similar shape over all \pt range 
for all small systems. A hint of suppression in central collisions and a 
hint of enhancement in peripheral collisions at high-\pt could be 
explained as events with high-\pt mesons having smaller underlying event 
multiplicity.

The \vphi meson production in the most central collisions shows a trend 
to less suppression than the \pio meson production at moderate \pt. 
However, the \rxa for both mesons are in agreement within uncertainties. 
This might suggest that strangeness-enhancement effects cannot be 
precluded.

Although the hot-nuclear-matter effects, such as strangeness enhancement 
and jet quenching, are imperceptible in small collision systems, \vphi 
meson \rxa in \pdheau collisions are in good agreement with the 
string-melting version of {\sc ampt} calculations, whereas the default 
version of {\sc ampt} calculations underpredict the data. Although 
{\sc pythia}/Angantyr and {\sc epps16} and {\sc ncteq15} nPDF 
calculations describe the experimental results within uncertainties, the 
predicted \rxa values do not describe measured \rxa ordering.

Experimental results in \pal collisions are better described with the 
default version of the {\sc ampt}-model calculations and are also consistent 
with {\sc pythia} model and nPDFs calculations. Hence, in spite of some 
collective effects observed in \pal collisions at \sqsntwo at 
backward rapidity, at midrapidity the QGP formation does not reveal 
itself.

The obtained results are in favor of the QGP formation in small 
collision systems. However, the volume and lifetime of the medium 
produced in these collisions might be insufficient for observing 
strangeness-enhancement and jet-quenching effects.

Comparisons with model predictions suggest, that \vphi-meson production 
in \pdheau collisions at \sqsntwo might be driven by mechanisms 
additional to nPDF. The hadronization process in \pal collisions could 
be interpreted within the frame of the fragmentation model and the 
influence of a coalescence mechanism seems to be negligible at 
midrapidity. The larger \pdheau systems can be well described by 
invoking the coalescence mechanism. Further studies of QGP effects in 
small collision systems and comparison of all available experimental 
results to the theoretical predictions, considering hot- and 
cold-nuclear-matter effects, are necessary for revealing the possibility 
of QGP formation. Particularly, the comparison of obtained \vphi meson 
results to $p(\bar{p})$ production in small collision systems at \sqsn = 
200 GeV at midrapidity can reveal a role of recombination or radial flow 
in observed \vphi and \pio \rxa ordering.



\begin{acknowledgments}

We thank the staff of the Collider-Accelerator and Physics
Departments at Brookhaven National Laboratory and the staff of
the other PHENIX participating institutions for their vital
contributions.  
We acknowledge support from the Office of Nuclear Physics in the
Office of Science of the Department of Energy,
the National Science Foundation,
Abilene Christian University Research Council,
Research Foundation of SUNY, and
Dean of the College of Arts and Sciences, Vanderbilt University
(USA),
Ministry of Education, Culture, Sports, Science, and Technology
and the Japan Society for the Promotion of Science (Japan),
Natural Science Foundation of China (People's Republic of China),
Croatian Science Foundation and
Ministry of Science and Education (Croatia),
Ministry of Education, Youth and Sports (Czech Republic),
Centre National de la Recherche Scientifique, Commissariat
{\`a} l'{\'E}nergie Atomique, and Institut National de Physique
Nucl{\'e}aire et de Physique des Particules (France),
J. Bolyai Research Scholarship, EFOP, the New National Excellence
Program ({\'U}NKP), NKFIH, and OTKA (Hungary),
Department of Atomic Energy and Department of Science and Technology
(India),
Israel Science Foundation (Israel),
Basic Science Research and SRC(CENuM) Programs through NRF
funded by the Ministry of Education and the Ministry of
Science and ICT (Korea),
Ministry of Education and Science, Russian Academy of Sciences,
Federal Agency of Atomic Energy (Russia),
VR and Wallenberg Foundation (Sweden),
University of Zambia, the Government of the Republic of Zambia (Zambia),
the U.S. Civilian Research and Development Foundation for the
Independent States of the Former Soviet Union,
the Hungarian American Enterprise Scholarship Fund,
the US-Hungarian Fulbright Foundation,
and the US-Israel Binational Science Foundation.

\end{acknowledgments}


%
 
\end{document}